# Robust intralayer antiferromagnetism and tricriticality in a van der Waals compound: VBr$_3$ case

Dávid Hovančík[1], Marie Kratochvílová[1], Tetiana Haidamak[1], Petr Doležal[1], Karel Carva[1], Anežka Bendová[1], Jan Prokleška[1], Petr Proschek[1], Martin Míšek[2], Denis I. Gorbunov[3], Jan Kotek[4], Vladimír Sechovský[1] and Jiří Pospíšil[1*]

[1] Charles University, Faculty of Mathematics and Physics, Department of Condensed Matter Physics, Ke Karlovu 5, 121 16 Prague 2, Czech Republic

[2] Institute of Physics, Czech Academy of Sciences, Na Slovance 2, 182 21 Prague 8, Czech Republic

[3] Hochfeld-Magnetlabor Dresden (HLD-EMFL), Helmholtz-Zentrum Dresden-Rossendorf, 01328 Dresden, Germany

[4] Department of Inorganic Chemistry, Faculty of Science, Charles University, Hlavova 8, 128 40 Prague 2, Czech Republic.

## ABSTRACT

We studied magnetic states and phase transitions in the van der Waals antiferromagnet VBr$_3$ experimentally by specific heat and magnetization measurements of single crystals in high magnetic fields and theoretically by the density functional theory calculations focused on exchange interactions. The magnetization behavior mimics Ising antiferromagnets with magnetic moments pointing out-of-plane due to strong uniaxial magnetocrystalline anisotropy. The out-of-plane magnetic field induces a spin-flip metamagnetic transition of first-order type at low temperatures, while at higher temperatures the transition becomes continuous. The first-order and continuous transition segments in the field-temperature phase diagram meet at a tricritical point. The magnetization response to the in-plane field manifests a continuous spin canting which is completed at the anisotropy field $\mu_0 H_{MA} \approx 27$ T. At higher fields the two magnetization curves above saturate at the same value of magnetic moment $\mu_{sat} \approx 1.2$ $\mu_B$/f.u., which is much smaller than the spin-only ($S = 1$) moment of the V$^{3+}$ ion. The reduced moment can be explained by the existence of an unquenched orbital magnetic moment antiparallel to the spin. The orbital moment is a key ingredient of a mechanism responsible for the observed large anisotropy. The exact energy evaluation of possible magnetic structures shows that the intralayer zigzag antiferromagnetic order is preferred which renders the antiferromagnetic ground state significantly more stable against the spin-flip transition than the other options. The calculations also predict that a minimal distortion of the Br ion sublattice causes a radical change of the orbital occupation in the ground state, connected with the formation of the orbital moment and the stability of magnetic order.

## I. INTRODUCTION

Van der Waals (vdW) magnets provide a natural platform for studying two-dimensional (2D) magnetism and its potential for advanced technologies like magnetooptics, spintronics, etc.[1-6]. Transition-metal trihalides, $TX_3$ ($T$ = V, Cr; $X$ = Cl, Br, I), form an important group of vdW magnets. VI$_3$[7-9], CrBr$_3$[10, 11], and CrI$_3$[12-15] become ferromagnetic (FM), whereas VBr$_3$[7, 16, 17], VCl$_3$[7], and CrCl$_3$[18, 19] are antiferromagnetic (AFM) at low temperatures. Historically, FM materials were considered to be more interesting from the point of view of applications, but recent discoveries show that AFM may hold even larger potential for magnetism-based information storage and spintronics[20-22]. The *TX*$_3$ compounds are dimorphic, adopting the rhombohedral BiI$_3$-type and the monoclinic AlCl$_3$-type (or related types) layered crystal structures[8, 9, 13, 14, 16, 23-25]. Both structure types are formed by stacking the *X-T-X* triple layers (further referred to as monolayers). The *T*-ion layer in the BiI$_3$-type structure has a graphenelike honeycomb form of a regular three-fold symmetry. The monolayers are equidistantly shifted, and the honeycomb network may be distorted in monoclinic structure[8, 26]. The weak *X-X* vdW bond between neighboring monolayers allows their easy separation. This unique property, combined with the magnetic ordering within a monolayer at finite temperatures, provides a testing base for 2D magnetism and offers promising opportunities to fabricate 2D nanoelectronic devices[27, 28].

The V trihalides undergo a structural phase transition from a high-temperature trigonal structure to the low-temperature monoclinic one at a temperature $T_s$ (≈ 79, 90, 97 K for VI$_3$, VBr$_3$, VCl$_3$)[16, 25]. The distortions of the $\beta$ angle in VBr$_3$ (90°→ 90.55°) and VI$_3$ (90°→ 90.45°) are only slightly different. Kong et al. reported VBr$_3$ antiferromagnetism below $T_N$ = 26.5 K with the magnetic moments perpendicular to the *ab*-plane (out of plane direction). We will mark this direction by the symbol *c\** to distinguish it from the *c*-axis, which is not perpendicular to the *ab* plane in the monoclinic structure. The recent publication[17] attributed the changes in Raman-scattering spectra around 90 K to the structural transition associated with a decrease in the crystal-structure symmetry from $R\bar{3}$ to C2/m. The authors also reported minor hysteresis loops observed in low magnetic fields and ascribed them to a canted AFM order.

Our work is mainly devoted to aspects of VBr$_3$ physics not treated in previous studies presented in the literature. The experiments were focused primarily on the influence of the magnetic field on the structural and magnetic phases in VBr$_3$, which can bring essential information on the mechanisms driving the specific phenomena. The structural-transition temperature $T_s$ was found intact by magnetic fields contrary to the reduction of $T_s$ of the isostructural ferromagnet VI$_3$ in the out-of-plane field (*H* ∥ *c\**)[24]. The observed dramatically different responses of $T_N$-related specific-heat anomalies and magnetization isotherms to the fields *H* ∥ *c\** and *H* ⊥ *c\** reveal strong uniaxial magnetocrystalline anisotropy. VBr$_3$, similar to other antiferromagnets, undergoes a magnetic-field-induced AFM → PM metamagnetic phase transition (MPT) at critical magnetic field $H_c$. We have observed a spin-flip transition in fields *H* ∥ *c\** that is typical for an Ising-like antiferromagnet. In fields *H* ⊥ *c\**, i.e. in the *a-b* plane, a continuous spin canting has been found running from the lowest fields and completed around $\mu_0 H_c$ ≈ 27 T, which can be considered a reasonable estimate of the anisotropy field. Such a high value of the anisotropy field indicates the presence of a significant V orbital magnetic moment. The existence of orbital moment can consistently explain also the size of the saturated magnetic moment of 1.2 $\mu_B$/f.u. observed in high magnetic fields ($\mu_0 H_c$ > 55 T). This value is much smaller than the spin-only value (2 $\mu_B$) of the magnetic moment expected for the V$^{3+}$ ion with a quenched orbital moment.

A closer inspection of the evolution of specific heat and magnetization isotherms in *H* ∥ *c\** reveals that the MPT at lower temperatures has a first-order transition character contrary to higher temperatures up to $T_N$, where the MPT becomes a continuous transition. This behavior is typical for antiferromagnets with a tricritical point that separates the first-order and the continuous metamagnetic transition segments in the field-temperature (*H-T*) magnetic phase diagram[29]. The archetype of this unique phenomenon is FeCl$_2$[29-31] also a layered vdW compound[32]. However, the tricriticality has not been reported in any other vdW antiferromagnet.

Previous first principles calculations for VBr$_3$ predicted a strong FM intralayer superexchange interaction like in VI$_3$[33]. This result suggests that the AFM structure should be composed of FM monolayers that are AFM coupled in the out-of-plane direction (layered AFM order), e.g., in the case of CrPS$_4$[34]. The calculation results excluded only the Néel AFM order inside layers caused by the AFM nearest-neighbor interactions. Nevertheless, sufficiently strong negative exchange interactions between distant V next nearest neighbors within the monolayer could also lead to more complex intralayer orders such as a stripe or a zigzag AFM order[35, 36]. Magnetism at a honeycomb lattice is typically described by interactions up to third nearest-neighbor ($J_1$-$J_2$-$J_3$ model). The zigzag AFM order has been identified in other quasi-2D systems on the honeycomb spin-lattice FePS$_3$[37, 38] and NiPS$_3$[39, 40]. The interlayer interaction was predicted to be an order of magnitude smaller than the intralayer one. Other studies were devoted only to the theoretical calculations of properties of VBr$_3$ monolayers[41, 42].

The current state of understanding of the physics of the vdW AFM VBr$_3$ motivated us to focus the present study on the role of crystal structure details on exchange interactions and the influence of the applied magnetic field on the magnetic states. To study the impact of crystal-structure details in determining the character of the exchange interactions and the magnetic ground state, we performed density functional theory (DFT) calculations. Spin-orbit coupling (SOC), which plays a vital role in the formation of the orbital moment, has been included. The calculations demonstrate that the ion network deformation is crucial for the stability of the AFM ordering of VBr$_3$. Our findings contradict the previous assumption of layered AFM ordering since we have found that zig-zag AFM order is energetically more favorable than FM order in layers. The predicted energy difference between this state and FM order is in agreement with the experimentally measured field needed to reorient spins. We have also shown that a particular displacement of Br atoms has a powerful impact on the ground state orbital occupation and exchange interactions.

## II. EXPERIMENTAL SECTION

### A. Material synthesis and single-crystal growth

The single crystals of VBr$_3$ have been grown from pure elements (V 99.9 %, Br$_2$ 99.5%) using the chemical vapor transport method. This approach prevented contamination of the final product by residuals from precursors readily used to produce Br$_2$, e.g. TeBr$_4$ used by Lyu et al.[17], in the reaction and transporting tube. First, a quartz tube with V metal powder kept at 300°C was evacuated overnight with simultaneous baking down to a vacuum of $10^{-7}$ mbar for proper degassing. Then, the tube was filled with 6N argon gas and cooled to -60°C by dipping into an ethanol bath cooled by dry ice. Subsequently, the stoichiometric volume of Br$_2$ liquid was injected inside the frozen tube, where it instantly solidified. The continuously cooled tube with the mixture of V powder and solid Br$_2$ was then pre-evacuated by a Scroll pump and evacuated by a turbomolecular pump for 5 minutes with no signatures of Br$_2$ evaporation. Finally, the sealed quartz tube was inserted into a gradient furnace where a thermal gradient of 460/350°C was kept for two weeks to transport all V metal from the hot part. The single crystals of the black-reflective color of several millimeter square dimensions have been obtained. The single crystals seem stable for several hours with no significant degradation effect. The desired 1:3 composition was confirmed by EDX analysis. The crystallinity and orientation of the single crystals were confirmed by the Laue method showing sharp reflections. The rhombohedral c-axis is perpendicular to the plane of plate-like crystals.

### B. Magnetization and specific heat study

Specific heat was measured by the relaxation method and magnetization data with a vibrating sample magnetometer in magnetic fields up to 14 T using a Quantum Design PPMS 14T (Quantum Design Inc.) and up to 18 T with a Cryogenic cryomagnet (Cryogenic Ltd.). The Néel temperature $T_N$ was determined from the temperature dependence of specific heat as the point of the balance of the entropy released at the phase transition. The field dependence of specific heat was measured point by point in a stable magnetic field. To probe the angular dependence of magnetization in *ac\**-plane and *ab*-plane (7 T) we used a homemade rotator for MPMS 7T with a rotation axis orthogonal to the applied magnetic field. The magnetization in pulsed magnetic fields up to ~58 T was measured at the Dresden High Magnetic Field Laboratory using a high-field magnetometer[43] with a coaxial pick-up coil system. Absolute values of the magnetization were calibrated using data obtained in steady fields. The measurements in magnetic fields were performed for two perpendicular directions of the field: in the *ab*-plane and perpendicular to the *a-b* plane. The *c*-axis in the monoclinic structure is not perpendicular to the *ab* plane. To avoid ambiguities, we use the symbol *c\** for the direction perpendicular to the *a-b* plane (*c\** ⊥ *ab* plane). In the trigonal structure *c\** is parallel to the *c*-axis (*c\** ∥ *c*). We extra note, that the single crystals are very thin and fragile plates moreover unstable in air, and commonly used solvents or glues[44]. It complicated the preparation and fixation of the samples for various experimental methods and caused slight uncertainty of their final masses which resulted in deviation of calculated absolute values of the physical quantities with 5-10%-error bar.

C. Theoretical calculations

Density functional theory (DFT) calculations employed the full-potential linear augmented plane wave (FP-LAPW) method, as implemented in the band structure program ELK[45]. Spin-orbit coupling (SOC) plays a crucial role in V trihalides. Therefore, it has been included in the calculations[46, 47]. The generalized gradient approximation (GGA) parametrized by Perdew-Burke- Ernzerhof[48] has been used to perform geometrical relaxation of the structure. We have used local density approximation (LDA) as the exchange-correlation potential to determine more subtle properties requiring high precision in energy, as are the exchange interactions and magnetic anisotropy energy since with GGA we have noticed numerical instabilities and convergence problems. These may be related to the presence of multiple local energy minima in the configurational space, or maybe of the same origin as those found in the pseudopotential-based calculation of vdW trisulfides[49]. Since the material is known to be a Mott insulator, we have included the effect of electron-electron correlations in terms of the Hubbard correction term $U = 4.3$ eV [50-52] acting on V 3d electrons. Double counting was treated in the fully localized limit. Similar DFT+U+SOC calculations have already successfully described the quasi-2D compound $VI_3$[52]. The entire Brillouin zone has been sampled by $10 \times 10 \times 5$ k-points and the convergence w.r.t. k-mesh density has been verified. For total energy calculations, increased angular momentum cut-off of the expansion into spherical harmonics has been used with $l_{max} = 14$ for the APW functions and $l_{maxo} = 8$ for the muffin-tin density and potential. To evaluate the interlayer interaction $J_L$ a unit cell doubled in the z-direction was used. Energies of calculated self-consistent ground states with forced FM and AFM interlayer alignment (LAFM) then allow us to calculate $J_L = E_{FM} - E_{LAFM}$ for different possible geometries, similar to the pressure dependence of $J_L$ calculated already for $VI_3$[51]. To compare the energies of the possible magnetic orderings inside the layer we double the unit cell in one of the planar directions. The basis with 4 V atoms in the layer allows us to evaluate energies of AFM Néel, stripe, and zig-zag orders, as well as FM order[53].

III. RESULTS AND DISCUSSION

Two transitions were detected in specific heat data (see Fig. 1a) in agreement with Kong et al.[16] and Lyu et al.[17]. The sharp peak at $T_s = 90$ K corresponds to the structural transition between the monoclinic and trigonal phases. The λ-shape anomaly at $T_N = 26.5$ K indicates a second-order phase transition between the low-temperature AFM state and the PM state (AFM ↔ PM).

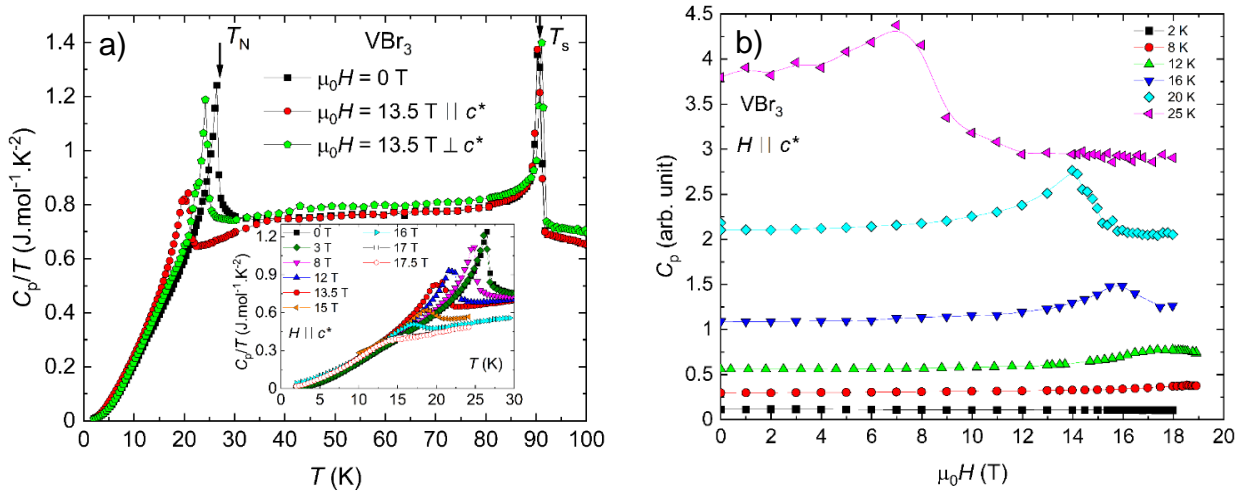

**FIG. 1.** a)The temperature dependence of specific heat of VBr$_3$ in zero magnetic field (■) and the field of 13.5 T applied parallel (●) and perpendicular (⬢) to $c^*$. The arrows mark the position of $T_N$ and $T_s$. The inset shows the variation of the $T_N$-related anomaly for the magnetic fields applied in the $c^*$ direction. b) The specific-heat isotherms of VBr$_3$ at selected temperatures for the magnetic field parallel to $c^*$.

The structural phase transition is intact by magnetic fields up to 18 T (13.5 T data are shown in Fig. 1a) applied in both principal directions. This result contrasts with the significant field dependence of the corresponding structural phase transformation of isostructural vdW FM VI$_3$ in fields parallel to $c^{*24}$, which can be understood as a result of the ferromagnetic correlations detected in Raman spectra[54], combined with strong magnetoelastic interaction. On the other hand, the applied magnetic field pushes the $T_N$-related anomaly in VBr$_3$ to lower temperatures and smears it out. These effects are much more pronounced for $H \parallel c^*$. No sign of magnetic phase transition is detected in magnetic fields > 17.5 T.

Specific-heat isotherms measured in varying magnetic fields $H \parallel c^*$ up to 18 T are shown in Fig. 1b. A broad-peak anomaly on the $C_p(H)$ isotherm manifests the second-order metamagnetic transition (AFM→PM) at temperatures from 12 K and $T_N$. For temperatures lower than 12 K, no anomaly is detected on the $C_p$ vs. $H$ plot. That is consistent with $C_p$ vs. $T$ data for $T < 12$ K in the inset of Fig. 1a which barely change with the applied magnetic field. Considering Maxwell's relation (eq 1)

$$\left(\frac{\partial S}{\partial H}\right)_T = \left(\frac{\partial M}{\partial T}\right)_H \quad (1)$$

this behavior indicates a negligible temperature dependence of magnetization in different magnetic fields $T < 12$ K.

The 2-K $M(H)$ isotherms $H \parallel c^*$ and $H \perp c^*$ in static fields up to 18 T, are shown in Fig. 2a. The $M(H)$ measured in $H \perp c^*$ remains linear up to 15 T and then becomes convex. The convexity is more pronounced with further increasing $H$. The increasing $H \parallel c^*$ field induces a steep S-shape increase in fields above 15.5 T with an inflection point at 16.9 T. This anomaly resembles a spin-flip transition typical for antiferromagnets with a strong uniaxial magneto-crystalline anisotropy. The observed field hysteresis $\mu_0 \Delta H \approx 0.3$ T suggests a first-order metamagnetic transition.

To reveal the limits of the AFM-phase stability in the $H$-$T$ phase space, we employed pulsed magnetic fields up to 58 T. The 1.9-K $M(H)$ isotherms are displayed in Fig. 2a. For $H \parallel c^*$, the S-shape profile is broader than in static fields, most likely due to a slower reaction of the magnetic moments to the fast field pulse. The magnetization in fields above the transition gradually saturates to the final value of 1.2 $\mu_B$/f.u. at 58 T. The $M(H)$ curve for $H \perp c^*$ is linear, becoming convex above 20 T. The convexity increases with increasing $H$ up to the inflection point at $\approx 23$ T, where the AFM → PM transition appears. Above 27 T, the magnetization gradually saturates and approaches the $H \parallel c^*$ curve. Both $M(H)$ curves join above 40 T. The observed saturated moment of 1.2 $\mu_B$/f.u. is much lower than that expected for the V$^{3+}$ spin-only magnetic moment (2 $\mu_B$/f.u.)[16, 41, 42, 47, 55-57]. It is worth noting that the experimentally observed saturated moment for VBr$_3$ compares well with that measured on the FM VI$_3$ single crystals in which a large V orbital magnetic moment has been recently confirmed by X-ray magnetic circular dichroism experiments[58]. This result, together with the observed strong magnetic anisotropy and large $H_c$, fields are clear indications that the V-ion in AFM VBr$_3$ also bears a significant orbital moment. The results of our anisotropy study and $H \perp c^*$ magnetization isotherms are shown in Supplemental Material (SM)[59].

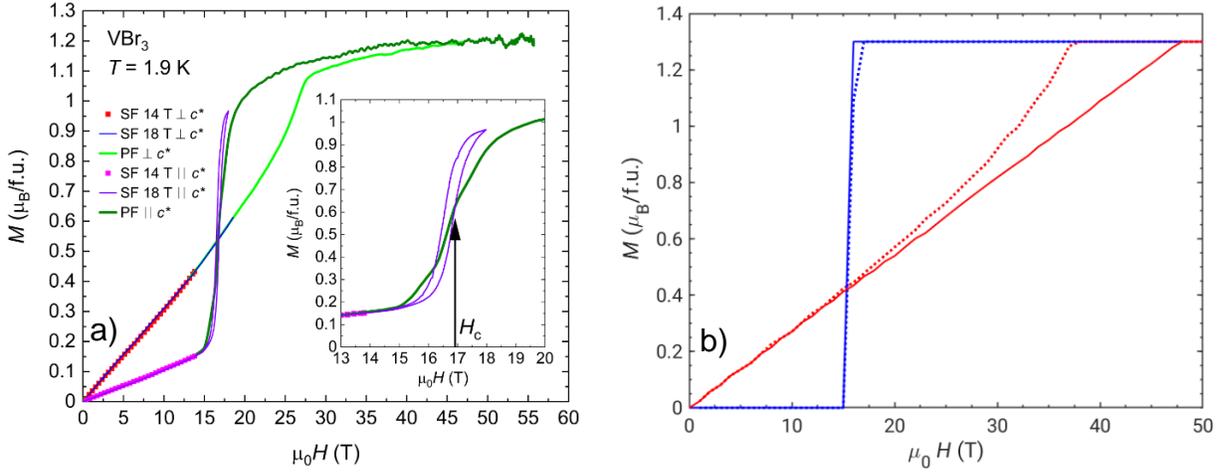

**FIG. 2.** a)The magnetization isotherms of VBr$_3$ measured at 1.9 K in static fields (SF) up to 18 T and pulsed fields (PF) $H \parallel c^*$ and $H \perp c^*$ up to 58 T. Inset: The detail of the plots for $H \parallel c^*$ between 13 and 20 T. The arrow points to the critical field $H_c$. The absolute value of the calibrated magnetization isotherms can vary with an error bar ± 10%. b) Stoner-Wohlfarth simulation. Plotted $M$ along $H$ per one site as a function of applied external field $H$, for $H$ parallel to the easy axis (blue) or perpendicular to the easy axis (red). Solid lines are calculated according to the model assuming purely uniaxial anisotropy ($K_2 = 0$ meV), while calculations depicted with dashed lines include a higher-order term ($K_2 = 0.2$ meV).

Magnetization isotherms simulations based on the Stoner-Wohlfahrt model for the two sublattices have recovered some characteristics of the observed $M(H)$ curves at zero temperature. This model assumes two macroscopic moments representing the two spin sublattices described by unit moments $m_1$ and $m_2$. These are coupled by an effective interaction $J^*$ and subjected to external field $H$ and magnetic anisotropy[60]. The corresponding free energy expressed per one V atom is then given by (eq 2)

$E = -J^* m_1 \cdot m_2 + E_{AN}(m_1, m_2) + H \cdot (m_1 + m_2) \mu_{sat}$ (2)

Assuming uniaxial anisotropy with an easy axis and magnitude $K_1$, we may rewrite the energies using the angles $(\vartheta_1, \vartheta_2)$ between the two "moments" and the easy axis (eq 3).

$E_{AN}(m_1, m_2) = K_1 \cdot (\sin^2 \vartheta_1 + \sin^2 \vartheta_2)$ (3)

The external magnetic field is applied in the direction given by unit vector $h$. We search for moments directions minimizing the free energy. The observable quantity is the projection of total magnetization to the direction of the field, denoted as $M$, per one V atom it is given as (eq 4) (value $\mu_s = 1.3$ $\mu_B$ was used)

$M = h \cdot (m_1 + m_2) \mu_s$ (4)

The fact that the $M(H)$ experimentally observed in the $H \parallel c^*$ regime almost reaches saturation after the sudden increase indicates that a spin-flip has occurred[61]. This behavior can be reproduced in this model if $K_1 \sim J^*/2$ or higher. Here we assumed $J^* = -1.2$ meV, $K_1 = 0.6$ meV, which provides good agreement with the experimentally observed dependence for $H \parallel c^*$. But for the $H \perp c^*$ curve a slower approach to saturation is predicted- see Fig. 2b. There can be many reasons for this, given the limitations of the model. As the next step, we have extended the model by including a higher-order anisotropy term $K_2 \cdot (\sin^4 \vartheta_1 + \sin^4 \vartheta_2)$ which is present in systems with hexagonal or rhombohedral symmetry[62]. These terms were found to strongly affect field-induced magnetization dynamics in antiferromagnets despite their small value[63]. A calculation assuming $K_2 = 0.2$ meV leads to markedly improved agreement with the observed slope for high $H \perp c^*$ (compare solid and dotted curves in Fig. S5 in SM[59]). This indicates that the specific behavior of $M(H)$ curves can be explained by the difference in the presence of anisotropy terms of fourth-power in $\sin \vartheta$.

The experimentally observed non-zero slope for small $H \parallel c^*$ could be ascribed to an inclination of the field from the easy axis, or the high content of defects in samples. Since the angular dependence of magnetization (Fig. S3 in SM) shows that the $c$ direction corresponds to the easy axis, the latter explanation is more plausible. We have tested the model for various values of $K_1$ and inclinations of field direction from the easy axis results are displayed in SM[59] Fig. S4. The employed semiclassical spin dynamics does not capture the effect of quantum fluctuations. However, a recent study performed also on the $J_1$-$J_2$-$J_3$ model for the honeycomb lattice finds that quantum fluctuations are strongly suppressed for the case of S = 1, so that the boundaries between different phases are only slightly changed as compared to the classical solution[64]. Therefore we do not expect it to affect the main conclusions drawn from the semiclassical approach. Quantum fluctuations in AFM V trihalides deserve further study.

Since there are indications of a significant orbital moment value in VBr$_3$, we have examined how it could affect the $M(H)$ curve. A crucial contribution to magnetic anisotropy originates from crystal field effects (magnetocrystalline anisotropy). This interacts with the orbital moment directly, but its effect is transferred to spin via the spin-orbit interaction (SOI). Therefore, a sizable orbital moment can justify the high anisotropy that we used in our model, although the total moment is predominantly of spin origin. Furthermore, we have considered it as an extension of our model where the spin and the orbital moment would be treated separately. However, within the expected values of the magnitude of spin-orbit interaction, this does not lead to a change of properties that would be observable. A detailed description is provided in SM.

Fig. 3 shows that the increasing temperature up to 12 K has a tiny effect on the first-order MPT (FOMPT) in the $M(H)$ curves in the field $H \parallel c^*$. The critical field at 12 K is $\mu_0 H_c$ = 16.4 T. This behavior reasonably matches the $C_p(H)$ dependence that is related via Maxwell's relation (see Eq. 1). The second-order (continuous) MPT (SOMPT) accompanied by magnetic fluctuations appears as a bump on the isothermal $C_p$ vs. $H$ plot while the anomaly disappears at FOMPT (Fig. 1b) because the released latent heat has vanished by the principle of the relaxation method in PPMS apparatus (see as the example in ref [65]). When inspecting Fig. 1b with decreasing temperature we observe a smeared bump on the $C_p$ vs. $H$ (a signature of SOMPT dependence) still at 12 K whereas a FOMPT at this temperature is suggested by magnetization behavior. We take this as a sign of TCP proximity.

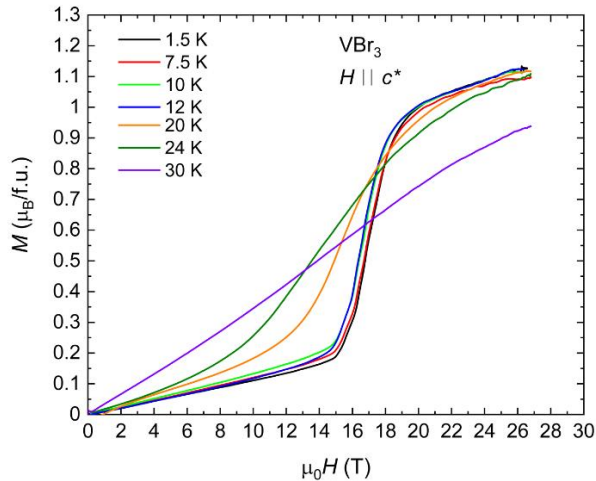

**FIG. 3.** The magnetization isotherms of VBr$_3$ measured at various temperatures in pulsed fields (PF) for $H \parallel c^*$ up to 27 T.

This evolution reflects the increasing influence of thermal fluctuations leading to the change of the character of the MPT to a continuous SOMPT at temperature interval $T_{cp} < T < T_N$. It indicates that VBr$_3$ belongs to the family of antiferromagnets in which the MPT is a FOMPT at low temperatures and in the highest magnetic fields whereas, at higher temperatures and lower fields, a continuous (second-order), i.e. SOMPT is observed. The SOMPT in the zero-field limit is corroborated by the λ-anomaly at $T_N$ in the temperature dependence of specific heat, in clear contrast to rare systems with first-order AFM transition with the peak-like anomaly[66-69]. The FOMPT is characterized by a sudden reversal of AFM-coupled FM sublattice(s) to the direction of the applied field. The high-field ($H > H_c$) state is then characterized by field-polarized magnetic moments. It resembles an FM alignment of magnetic moments, however, it is a paramagnetic (not ferromagnetic) state[29]. It is used to be called a polarized paramagnet (PPM) regime[70].

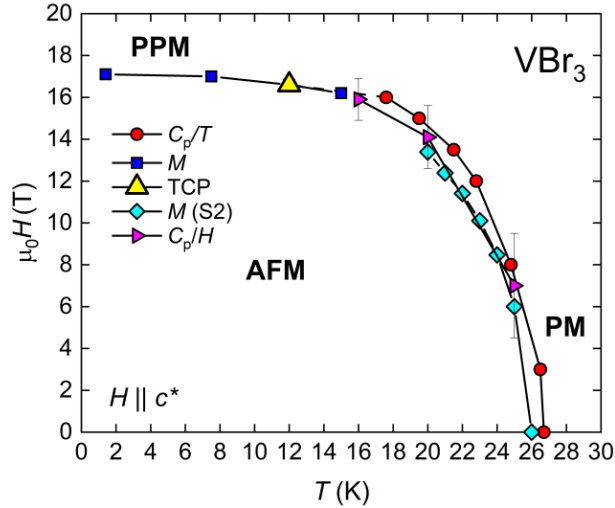

**FIG. 4.** The *H-T* phase diagram of VBr$_3$ for magnetic field applied along the *c**. The Blue curve represents the inflection points on MPT in pulsed field data, the red curve the position of the anomaly in specific heat data ($C_p$), the cyan curve represents the kink of MPT in steady field data (see Fig. S1), and magenta point shows the position of anomalies in field dependence of specific heat data. The yellow point shows the estimated position of TCP. The TCP is tentatively placed at $T_{TCP}$ = 12 K, $\mu_0 H_{TCP}$ = 16.4 T. *M*(S2) represents data measured on a sample of lower quality (less stable in air) for comparison.

The Ising antiferromagnet FeCl$_2$[30, 31] with competing FM and AFM exchange interactions is considered the archetype of this interesting family of materials. To retain the FOMPT above 0 K and to permit the occurrence of hysteresis at MT, a ferromagnetic intra-sublattice exchange is necessary for a simple Ising system. The FM exchange in VBr$_3$ is documented by the Heisenberg exchange interactions in the third nearest-neighbor model-see Table 2 where $J_1 > 0$. In such a case a ratio $\tau^* = T_{TCP}/T_N \approx 0.45$ for VBr$_3$ is established and proportional to (eq 5)

$\tau^* = 1 - (A/3\Gamma)$    (5)

where $A$ and $\Gamma$ are molecular field coefficients[30].

The *H-T* phase diagram of VBr$_3$ for $H \parallel c^*$ is displayed in Fig. 4. The PM ↔ AFM phase-transition line has two parts: a low-temperature part of FOMPTs and has high-temperature part of SOMPTs separated by the tricritical point (TCP)[29, 71] which we tentatively place at [12 K, 16.4 T] in the *H-T* phase diagram. Unfortunately, we do have not more *M* vs. *H* and $C_p$ vs. *H* isotherms data measured at steady fields enabling us to determine $T_{TCP}$ and $H_{TCP}$ coordinates more precisely. The lack of experimental facilities providing steady fields within a reasonable *T-H* space prevents performing magnetization measurements for a reliable study of critical coefficients in the interesting case of 2D antiferromagnet with a TCP.

We emphasize that VBr$_3$ is not a single non-Ising antiferromagnet with FOMPTs and SOMPTs segments in the *H-T* phase space separated by TCP. Analogous behavior is found also in some antiferromagnets characterized by strong uniaxial anisotropy[71-73] as well as in one exhibiting strong orthorhombic anisotropy[74, 75].

We have used first-principles calculation methods to evaluate the energies of the FM-ordered system, 3 plausible intralayer AFM orderings, and the layered AFM state. First calculations were performed for the lattice geometry determined by X-ray diffraction at 100 K, where the lattice parameters were found to be $a$ = 6.3711 Å, $c$ = 18.3763 Å, and the Br planes are placed at $h_{Br}$ = 0.07928 $c$ above (or below) the V planes[16]. Our calculations predict that the magnetic ground state has a zigzag AFM (Fig. 5a) order and not the suggested layered AFM structure consisting of ferromagnetically ordered layers with antiparallel orientation between neighboring layers[33] (Fig. 5b).

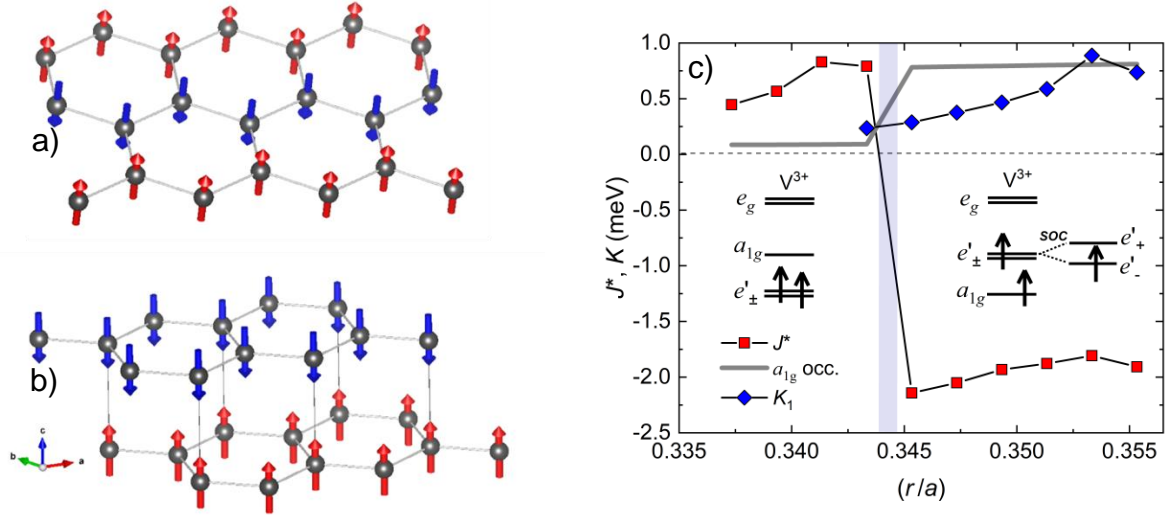

FIG. 5. a) The predicted AFM magnetic structure of $VBr_3$. Only a single layer of $V^{3+}$ ions is displayed. The AFM structure consists of FM zig-zag chains coupled antiferromagnetically within the plane. b) Layered AFM structure (ferromagnetically ordered layers with antiparallel orientation between neighboring layers)[33]. Two layers of $V^{3+}$ ions are displayed, and lines in the vertical direction connect V ions stacked on top of each other (differing only in the $z$ coordinate). c) Calculated effective exchange $J^*$ and anisotropy $K_1$ as a function of the distance $r$ (in multiples of lattice parameter $a$). The occupation of the $V^{3+}$ $d$-states is calculated and schematically displayed in the diagrams.

All other considered AFM orders, including the layered AFM state, are energetically less favorable (Table 1). Using calculated energies of magnetic configurations one can map the problem to an effective Heisenberg Hamiltonian (eq 6),

$$H = -\frac{1}{2}\Sigma_{i,j} J_{ij}\, s_i \cdot s_j \qquad (6)$$

where $s_i$ is the unit vector corresponding to the $i$-th spin in the system, $J_{ij}$ is the exchange energy between the $i$-th and $j$-th spins, and the sums run over magnetic atoms in the system. Magnetism at a honeycomb lattice is efficiently described by the Heisenberg model with interactions up to the third nearest neighbor ($J_1$-$J_2$-$J_3$ model, schematically depicted in Fig. 6b). Individual exchange interactions can be calculated from the calculated four energies of the four plausible magnetic order[76], and are shown in Table 2. The energy difference between the AFM zigzag state and the FM state corresponds to the stability of the system w.r.t. a field-induced spin-flip transition. It can be denoted as $2\cdot J^*$, an effective interaction between the two AF coupled sublattices (in $J_1$-$J_2$-$J_3$ model it equals $-J_1 - 4J_2 - 3J_3$). We find $J^* = -1.9$ meV per V atom. This finding appears to be consistent with the experimental knowledge about the presence of the AFM state and its stability, as the calculated $J^*$ is rather close to the value we could use to simulate the experimentally observed spin reorientation transition (Fig. 2) within the Stoner-Wohlfarth model.

**Table 1:** The calculated energies of magnetic structures relative to the zigzag AFM.

| Type of magnetic structure | Relative total energies $E$ (meV/f.u.) | |
|---|---|---|
| | 100 K structure | Relaxed structure |
| AFM zigzag | 0 | 0 |
| FM | +1.91 | +2.66 |
| Layered AFM | +1.65 | +2.09 |
| AFM stripe | +3.26 | +4.87 |
| AFM Neél | +6.04 | +4.81 |

**Table 2:** The calculated Heisenberg exchange interactions $J_i$ (in meV)

| Interaction | $J_1$ | $J_2$ | $J_3$ |
|---|---|---|---|
| 100 K structure | 1.87 | -0.57 | -0.46 |
| Relaxed structure | 1.76 | -0.33 | -1.04 |

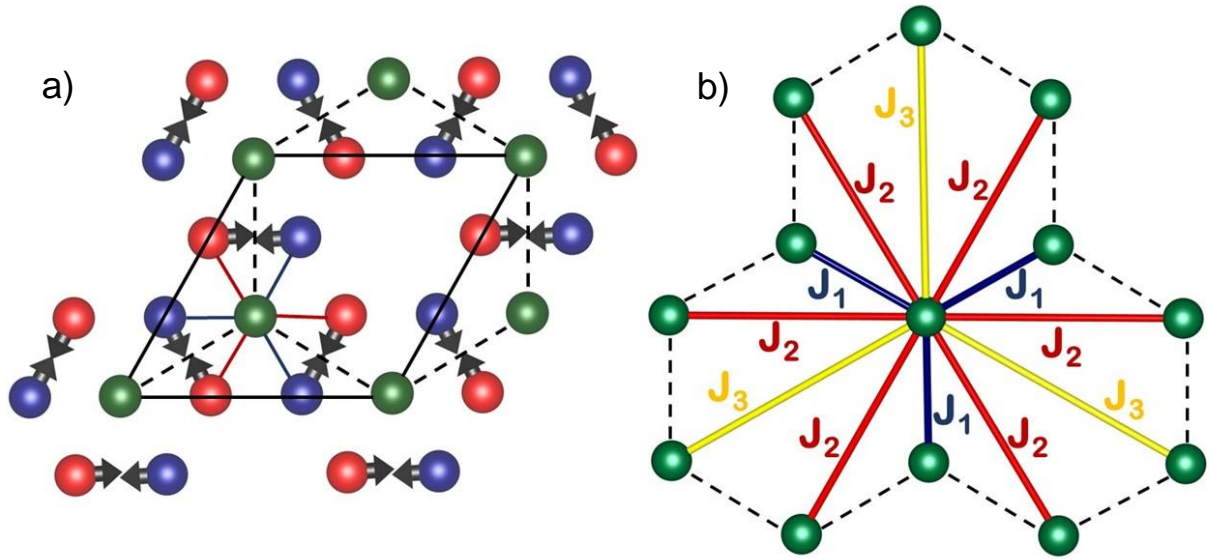

**FIG. 6.** a) VBr$_3$ lattice plane, predicted relaxation is shown. b) Heisenberg model with interactions up to the third nearest neighbor ($J_1$-$J_2$-$J_3$).

Note that the magnetically ordered state is present only in the low-temperature structure below $T_s$. The low-temperature structure is only known to be slightly distorted so that its symmetry is reduced to the monoclinic[22], but the structure details about anion positions are unknown. Therefore, as the next step, we performed geometrical relaxation of the atomic positions, starting from the monoclinically distorted structure. The optimized structure exhibits the following significant changes compared to the original one: i) a decrease of Br plane height $h_{Br}$ by only $\approx 0.002c$, ii) a slight planar shift of Br sites in the direction towards the midpoint between its two nearest V atoms, so that its distance $r$ from the hollow site in the honeycomb lattice increases from the value of $r_m = 0.349a$ (measured at 100 K) to $r_{opt} = 0.353a$ (see

Fig. 6a). A similar distortion has been predicted by first principles computational optimization in $BiI_3$, but in the opposite direction, anions have moved towards the vacant point in the cation honeycomb lattice[77].

We have studied the effect of these deformations on the difference between FM and zig-zag AFM order energies, $J^*$. Changing the Br height $h_{Br}$ does not introduce a significant change in the calculated exchange interactions to the extent suggested by the relaxation. The dependence of the effective exchange interactions $J^*$ on the planar Br distortion (Fig. 6a) is shown in Fig. 5c. The relaxed $h_{Br}$ was included in these calculations. Note that the exchange varies only slowly with $r$ (within numerical precision), while at a specific distance, a sudden change of the effective magnetic exchange occurs.

Like $VI_3$, the electronic structure of $VBr_3$ may converge to two strikingly different solutions: a state with a quenched orbital moment, typical for 3d transition metals in a medium-strength crystal field, or a state with a high orbital moment. This problem is connected with the position of different energy levels of trigonal symmetry electronic $d$ orbitals in the ground state, in particular, with the question of whether the $a_{1g}$ orbital or one of the $e'_g$ orbitals will be positioned above the Fermi level, as debated for $VI_3$[46, 47, 58, 78]. If the $a_{1g}$ orbital is unoccupied, $e'_g$ has to be fully occupied and the orbital moment would be suppressed. For the spin moment, the calculations predict values close to 2 $\mu_B$/f.u. in agreement with Hund's rules, but the experimental magnetic moment is 1.2 $\mu_B$/f.u. This sharp transition is associated with a change in the occupation of different electronic orbitals in the ground state. For $r < r_{crit}$ states with fully occupied $e'_g$ are preferred, while a state with $a_{1g}$ occupied turns out to be favorable for $r$ above this threshold value $r_{crit} = 0.344a$.

A correct Br position has to be used to obtain the correct ground state in calculations. For $r = r_{opt}$ our calculations predict a sizeable magnetocrystalline anisotropy 0.89 meV, a value that is reasonably close to that we have used to describe the field-induced spin-flip within the Stoner-Wohlfahrt model. For example, the assumption of the so-called ideal position (with $r = a/3$)[77] would lead to a different magnetic order as well as an electronic structure. A compression of $r$ by only ~3% from its optimal value is sufficient to overcome $r_{crit}$ and reach this different state. Our calculations also predict easy plane preference for that situation, together with significant orbital reoccupation for moment orientation in-plane. The state predicted for $r < r_{crit}$ does not correspond to current observations, but it could be reached by the application of pressure or if specific phononic modes would be significantly occupied.

A zigzag AFM magnetic structure was also predicted in vdW AFM $FePS_3$. The $VBr_3$ magnetization loops significantly differ from those of $FePS_3$ where specific magnetization plateaus (cascades) were detected[53, 79]. These occur due to a special combination of exchange interaction values that favor a situation with partially flipped moments within a small range of applied fields, where for example six moments in a unit cell point in one direction and two moments in the opposite direction[79]. Only a simple spin-flip transition was found in $VBr_3$, therefore, based on the suggested models, we suppose the simple out-of-plane zigzag intralayer AFM structure as displayed in Fig. 5a. without preference for intermediate partially flipped states.

## IV. CONCLUSIONS

We have grown high-quality single crystals of $VBr_3$ by chemical vapor transport directly from pure elements. The study addresses the character of antiferromagnetism and phase transitions by measuring specific heat and magnetization in static magnetic fields up to 18.5 T and pulsed fields up to 58 T. The structural transition at $T_s = 90$ K remains intact by magnetic fields, which is in contrast with the decrease of $T_s$ in the isostructural ferromagnet $VI_3$. $T_N$ was found to decrease with increasing magnetic field much faster for the out-of-plane field than for the in-plane-direction field. The magnetization response to the magnetic field is strongly anisotropic. A first-order metamagnetic spin-flip transition to PPM occurs at $\mu_0 H_c = 16.9$ T in the out-of-plane field at 2 K. This transition remains first-order at temperatures up to 12 K. At higher temperatures up to $T_N$ the AFM → PM SOMPT occurs at lower fields. These phenomena are observed in some Ising antiferromagnets with strong uniaxial anisotropy and competing AFM and FM interactions. Results of our study suggest $VBr_3$ to be a member of this group of materials with TCP at

≈ [12 K, 16.4 T]. Further experiments that we do have not available are needed to determine the coordinates of TCP precisely and determine the dimensionality of the system in terms of critical coefficients. In particular, high steady field (at least up to 20 T) magnetization and electrical transport measurements are desirable. To our best knowledge, VBr$_3$ seems to be the only vdW antiferromagnet besides the archetype FeCl$_2$ in which the tricriticality has been reported until now.

The in-plane magnetization curve represents continuous spin canting towards the PM state with spin moments fully oriented along the applied field in fields above 27 T (at 2 K). The saturated magnetic moment observed in fields above 50 T $\mu_{sat} \approx 1.2$ $\mu_B$/f.u. is much smaller than the spin-only moment (2 $\mu_B$) of a V$^{3+}$ ion. This indicates the existence of a significant orbital magnetic moment, which partly compensates for the spin moment similar to the VI$_3$ case.

Our calculations predict that the magnetic structure in VBr$_3$ is based on the intralayer antiferromagnetic order in the form of a zigzag pattern. We have also found that the relaxation of Br atomic position plays an important role in the electronic structure calculations and the resulting magnetic properties of the system. A small change in the distances of Br atoms from the hollow site in the honeycomb lattice leads to a sudden redistribution of orbital occupation in the ground state. This suggests that phonon modes leading to a similar displacement would be strongly coupled to magnetic order here.

## ASSOCIATED CONTENT

**Suplemental material**. Results of magnetocrystalline anisotropy study of VBr$_3$ compound by angular dependence of magnetization; extended results of specific heat, steady, and pulsed-field magnetization study; extra theoretical calculations supporting the conclusions concerning the magnetic structure of VBr$_3$.

## ACKNOWLEDGMENT

This work is a part of the research project GAČR 21-06083S which is financed by the Czech Science Foundation and project GAUK 938220 financed by the Charles University Grant Agency. The single-crystal growth and characterization, and experiments in steady magnetic fields were carried out in the Materials Growth and Measurement Laboratory MGML (see: http://mgml.eu) which is supported within the program of Czech Research Infrastructures (project no. LM2023065). This project was also supported by OP VVV project MATFUN under Grant No. CZ.02.1.01/0.0/0.0/15_003/0000487. We acknowledge the support from HLD at HZDR, a member of the European Magnetic Field Laboratory (EMFL) where the measurements in pulsed magnetic fields were done. This work was supported by the Ministry of Education, Youth and Sports of the Czech Republic through the e-INFRA CZ (ID: 90140). We appreciate fruitful discussions with K. Výborný. The authors are also indebted to Dr. Ross H. Colman for critical reading and correcting of the manuscript.

# Robust intralayer antiferromagnetism and tricriticality in a van der Waals compound: VBr₃ case

Dávid Hovančík[1], Marie Kratochvílová[1], Tetiana Haidamak[1], Petr Doležal[1], Karel Carva[1], Anežka Bendová[1], Jan Prokleška[1], Petr Proschek[1], Martin Míšek[2], Denis I. Gorbunov[3], Jan Kotek[4], Vladimír Sechovský[1] and Jiří Pospíšil[1*]

[1] Charles University, Faculty of Mathematics and Physics, Department of Condensed Matter Physics, Ke Karlovu 5, 121 16 Prague 2, Czech Republic

[2] Institute of Physics, Czech Academy of Sciences, Na Slovance 2, 182 21 Prague 8, Czech Republic

[3] Hochfeld-Magnetlabor Dresden (HLD-EMFL), Helmholtz-Zentrum Dresden-Rossendorf, 01328 Dresden, Germany

[4] Department of Inorganic Chemistry, Faculty of Science, Charles University, Hlavova 8, 128 40 Prague 2, Czech Republic.

**Results**

The steady field 14 T has shown only a tiny effect on linearly growing magnetization up to the temperature of 12 K (Fig. S1). At higher temperatures, a signature of an upturn on magnetization was detected near-maximal values of magnetic fields signalizing the presence of MPT which finally appears at a temperature just below $T_N$. The effect of the magnetic field perpendicular to the $c^*$-axis is very weak in agreement with the specific heat data. The steady field 14 T magnetization isotherms are linear with no sign of any effect (inset of Fig. S2). Only a very weak deviation from the linear trend appears approaching $T_N$. The data demonstrates pronounced magneto-crystalline anisotropy. These results agree with the work of Kong et al.[1]. The pulsed field data up to 58 T for $H \perp c^*$ (Fig. S2) shows the magnetic moment canting transition, which is very robust to the increasing temperature. The first more distinguished effect on the shape of the isotherm is detectable at 20 K in the vicinity of the $T_N$.

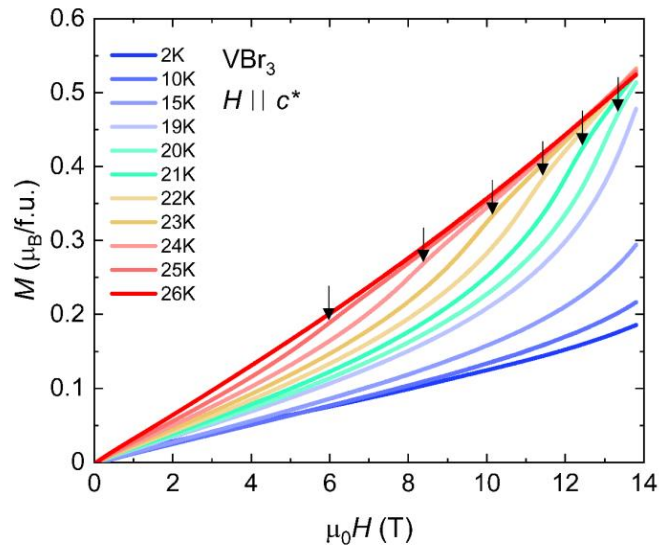

FIG. S1. The magnetization isotherms of VBr$_3$ measured with PPMS14T in the field applied $H \parallel c^*$. The black arrows mark the position of MPT, the values were used for the construction of the $H$-$T$ phase diagram in Fig. 4.

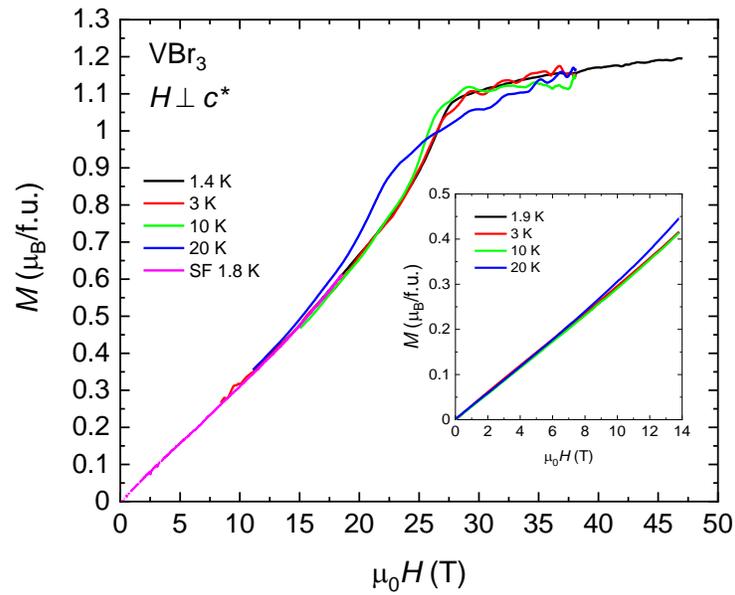

FIG. S2. The magnetization isotherms of VBr$_3$ measured at various temperatures in pulsed fields (PF) and static fields (SF) $H \perp c^*$ up to 47 T. Inset: The magnetization isotherms of VBr$_3$ measured at various temperatures in static fields $H \perp c^*$ up to 14 T without special orientation within the $ab$ plane.

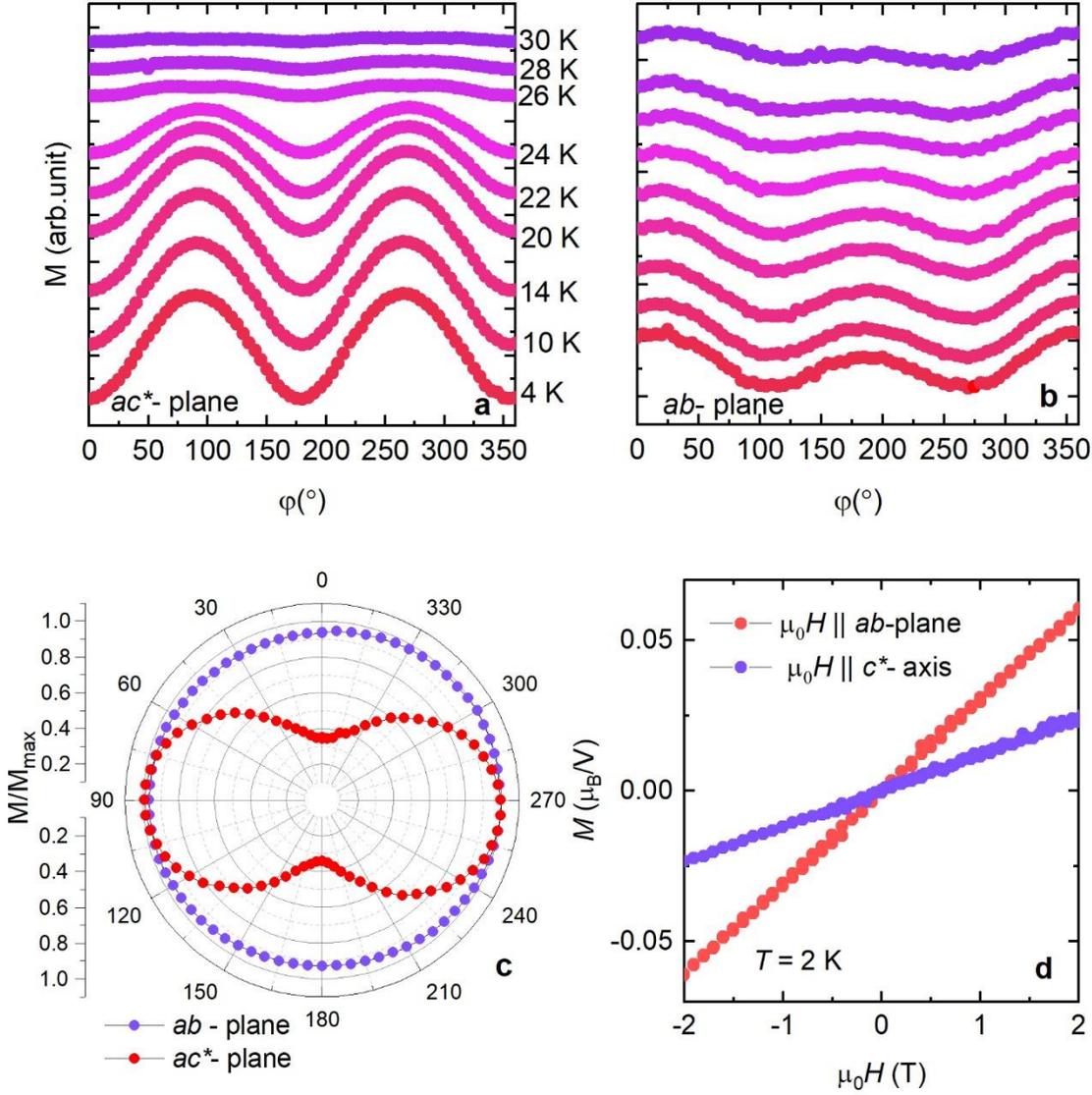

FIG. S3. Angular-dependent magnetization was taken in a) $ac^*$-plane (the $c^*$-axis corresponds to 0º and 180º) b) $ab$-plane for different temperatures measured in the magnetic field of 7 T. c) Polar plot of the rotational magnetization for $ab$-plane (blue) and $ac^*$-plane (red) taken at 4 K. d) Low-field magnetization isotherms measured along the $c^*$-axis (blue) and $ab$-plane (red).

To probe the magneto-crystalline anisotropy in VBr$_3$, we measured the angular dependence of the magnetization in $\mu_0 H = 7$ T applied in the $ac^*$ and $ab$ planes (see Fig. S3). The diamagnetic contribution of the rotator was subtracted from the data for both rotation planes. Below $T_N$, the magnetic signal in the $ac^*$ plane has two-fold symmetry with the minimum (maximum) along the $c^*$-axis ($ab$-plane), which corroborates the AFM state with the Néel vector aligned along the $c^*$-axis. The $ab$-plane magnetic signal also exhibits a weak two-fold-like symmetry but with a much smaller maximum/minimum signal ratio (almost isotropic) than the result in the $ac^*$ plane. That can indicate the slightly tilted Néel vector from the $c^*$-axis or the biaxiality of AFM anisotropy[2]. One can notice that a weak two-fold-like signal is already present just above $T_N$ for both planes, however, it is more apparent for the $ab$-plane given the small scale of the relative change of magnetization. The difference between the relative change of magnetic signal in $ac$-plane and $ab$-plane (almost isotropic) is clearly visible from the polar plot in Fig. S3c. The VBr$_3$ anisotropy and magnetic structure are in obvious contrast to the features of neighbor FM VI$_3$[3,4]. We have not traced any hysteresis loop in the low magnetic field along the $c^*$-axis (see Fig. S3), contrary to results reported and ascribed to the canted AFM easy-axis by Lyu et al.[5].

## Results – Stoner-Wohlfarth simulations

The two-sublattice Stoner-Wohlfarth model can provide an approximate picture of field-induced spin-flop and spin-flip transitions in antiferromagnets[6]. In Fig. S4a we show how the type of transition can be controlled by magnetic anisotropy (other parameters are the same as used in the main article text). For small fields applied along the easy axis, spins on the two sublattices are oriented antiparallel to each other due to the exchange interaction. For the most common case of small anisotropy (here we assume $K_1 = 0.1$ meV), at a sufficiently strong field, it is energetically favorable to rotate the moments into the spin-flop state so that there appears a projection into field direction, and the exchange, Zeeman and anisotropy energy are partially satisfied. With increased anisotropy, this transition occurs at a higher field with a smaller deviation from the easy axis, until a critical value of anisotropy is reached when it is more favorable to completely break the exchange interaction rather than the anisotropy energy, and a spin-flip occurs. We show the case with anisotropy slightly below this critical value ($K_1 = 0.4$ meV), so that small tilting from the easy axis occurs for a short range of $H$ values and a higher value ($K_1 = 0.6$ meV) with a complete spin-flip. Fields applied perpendicular to the easy axis lead to tilting of moments towards the field direction that is linearly dependent on the field magnitude. The field needed for saturation (complete tilt of moments) increases with increasing anisotropy.

We also examine how the results would be affected by possible deviation of the field direction from the orientation either along ($\vartheta_H = 0°$) or perpendicular ($\vartheta_H = 90°$) to the easy axis. The cases with field canted by 20° from these major directions (Fig. S4b)) have some features common with the experimentally observed $M(H)$ dependencies, but the deviation of 20° appears too large to be possible **(?)**. On the other hand, for misalignment less than 10° the change is hardly visible within the given range of the graph.

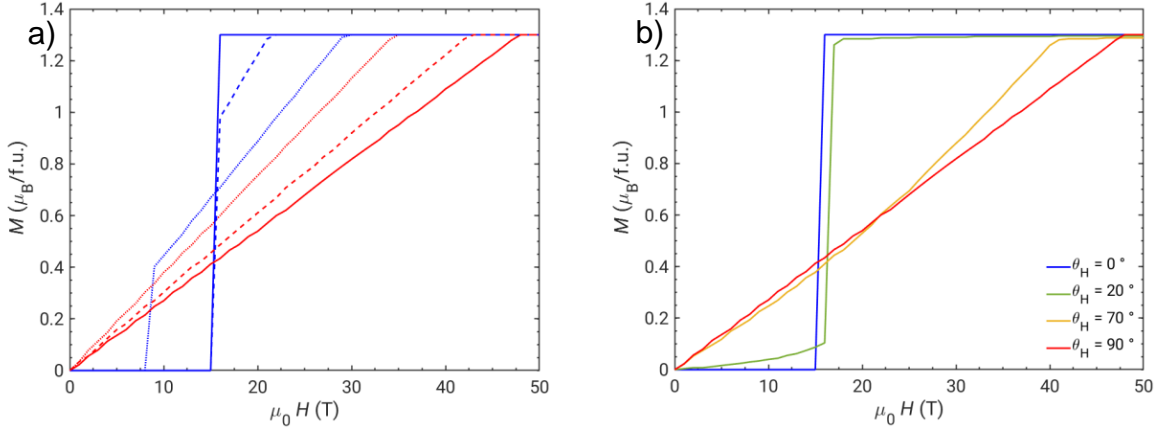

FIG. S4. Stoner-Wohlfarth simulations. Magnetization $M$ per f.u. as a function of the applied external field $H$: a) for different values of anisotropy $K_1$ with field applied either parallel to the easy axis (blue lines) or perpendicular to the easy axis (red lines). Dotted lines: $K_1 = 0.1$ meV. Solid lines: $K_1 = 0.53$ meV. Dashed lines: $K_1 = 1$ meV. b) for $H$ applied at different angles from the easy axis (with $K_1 = 0.53$ meV).

Due to the large size orbital moment and the relatively modest spin-orbit coupling for 3d electrons, we next consider a model where spin and orbital moments are treated separately, described by their own unit vectors $s_{1,2}$ and $l_{1,2}$. The moments are then given by the multiplication of these unit vectors with moment magnitudes ($\mu_s$ and $\mu_l$), they interact with the external field, and are coupled by SOI. Exchange interaction affects only the spin component. Concerning anisotropy energy, we assume that the magnetocrystalline part is dominant, and therefore affects only the orbital component. The free energy is then modified as follows (eq 1):

$$E = -J^* s_1 \cdot s_2 + E_{AN}(l_1, l_2) + H \cdot (s_1 + s_2)\mu_s + H \cdot (l_1 + l_2)\mu_l + \lambda(l_1 \cdot s_1 + l_2 \cdot s_2) \quad (1)$$

V ion is expected to be in the high spin state with spin moment of 2 $\mu_B$[7] and ab initio calculations predict it to be rather close to this value, between 1.85 and 2.03 $\mu_B$[1, 8-12]. Therefore we set the spin moment size to 2 $\mu_B$ and the orbital moment size to $\mu_L = 0.7$ $\mu_B$. SOI prefers antiparallel orientation to the spin so that the total moment is again 1.3 $\mu_B$. In the limit of very strong SOI, these moments remain rigidly coupled and behave like one moment, leading to a similar result as in the previous case. Assuming a smaller SOI ($\lambda = 3.5$ meV) we notice slight changes mainly in the $H \perp c^*$ curve (Fig. S5), as the variation of the angle between $s_i$ and $l_i$ provides an additional degree of freedom. That lowers total anisotropy energy connected with spins pointing perpendicular to the easy axis. The increase of $M$ with $H$ is then faster in this case. While the separation of spin and orbital moment within the classical model is disputable, it seems to improve slightly the agreement with the experiment.

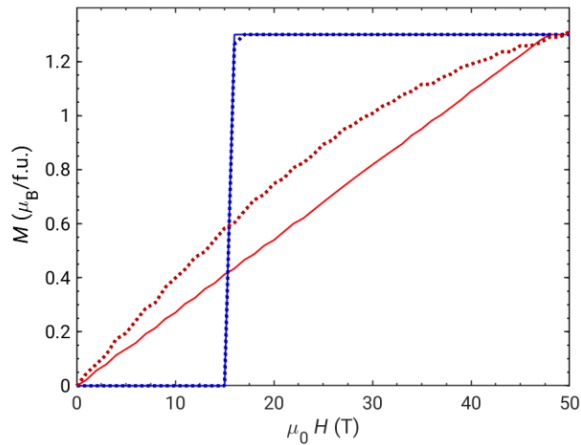

**FIG. S5.** Stoner-Wohlfahrt simulation. Plotted $M$ along $H$ per one site as a function of applied external field $H$, for $H$ parallel to the easy axis (blue) or perpendicular to the easy axis (red). Solid lines are calculated according to the model in (eq 4), while dashed lines come from the model with a separate orbital moment (eq 5).

Fig. S6 provides a detailed view of the dynamics in the model with separate spin and orbital contributions. For small fields applied along the easy axis, spins on one of the sublattices are oriented along the field, on the other one oppositely, similarly to the standard model with one moment per sublattice (studied above). The smaller orbital moments are oriented oppositely w.r.t the corresponding spins. Hence the total moment on one sublattice is ~1.3 $\mu_B$, and the total moment projected to the field direction is zero. A sufficiently strong field overcomes the effect of the exchange interaction and the second sublattice is flipped to be parallel to the field too. Orbital moments remain antiparallel to spins within the whole studied field range. Therefore the dynamics is almost identical to the simpler model considering one moment per sublattice (Fig. 3b of the main text[13]). For the field applied perpendicular to the easy axis, the behavior of spins is again similar to the case study above. However, since the anisotropy is acting on the orbital momentum, we see that it is advantageous for the system to tilt the spin momentum in the field direction slightly more than the orbital momentum. Some energy is thus gained from Zeeman energy on behalf of the spin-orbit contribution. The system effectively behaves similarly to the total moment system, but with slightly enlarged momentum, as shown by the red dotted curve in Fig. S6. We are not aware of any work discussing the possible separation of spin and orbital contributions in strong fields. It should be pointed out that this topic deserves further study.

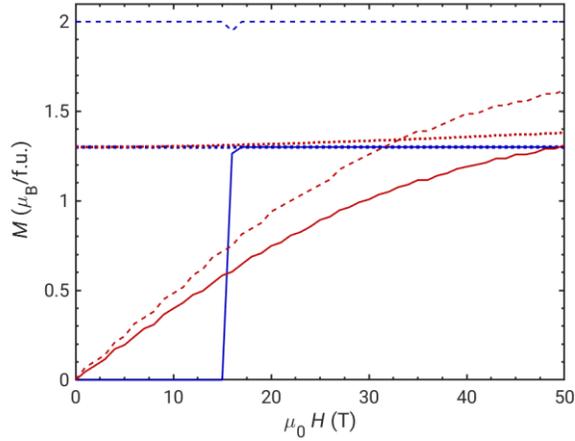

FIG. S6. Stoner-Wohlfarth simulations. Magnetic contributions of different sublattices (per f.u.) as a function of the applied external field $H$, for $H$ parallel to the easy axis (blue lines) or perpendicular to the easy axis (red lines). Solid lines: Total magnetization $M$. Dashed lines: spin contribution on sublattice 1 (projected to the direction of applied field $\mathbf{h}$ : $\mu_s \mathbf{h} \cdot \mathbf{s}_1$ ). Dotted lines: total magnetic momentum magnitude on sublattice 1 ( $|\mu_s \mathbf{s}_1 + \mu_l \mathbf{l}_1|$ ).